\documentstyle[amssymb,prl,aps,epsf]{revtex}
\begin{document}
\draft

\twocolumn[\hsize\textwidth\columnwidth\hsize\csname
@twocolumnfalse\endcsname
\title{Crossovers Between Elastic, Plastic and Quantum Creep in 2D YBa$_{2}$Cu$_{3}$O$_{7}$/PrBa$_{2}$Cu$_{3}$O$_{7}$ Superlattices}
\author{X.G. Qiu\cite{byline}, V.V. Moshchalkov, and Y. Bruynseraede}
\address{Laboratorium voor Vaste-Stoffysica en Magnetisme, Katholieke Universiteit Leuven, Celestijnenlaan 200D, B-3001 Leuven, Belgium}
\author{G. Jakob and H. Adrian}
\address{Institut f\"{u}r Physik, Johnannes Gutenberg Universit\"{a}t Mainz, Staudinger Weg 7, 55099 Mainz, Germany}
\maketitle

\begin{abstract}
Two-dimensional (2D) vortex dynamics was studied in an YBa$_{2}$Cu$_{3}$O$_{7}$/ PrBa$_{2}$Cu$_{3}$O$_{7}$ superlattice by measuring the I-V characteristics. In the high current limit, 2D collective creep was observed
with an activation energy characterized by $U(j)\propto j^{-\mu }$. A
dislocation mediated vortex melting happened when the temperature increased.
In the low current limit, the exponential growth of energy barrier for
elastic motion was prohibited by the plastic deformation of vortices. A
plateau in the resistive transition was observed, which was attributed to
possible quantum tunneling of vortices. Our results suggest that a 2D vortex
glass can not exist at any temperature, including T=0 K.
\end{abstract}
\vspace{-20pt}
\vskip2pc] \vskip2pc \narrowtext

The unusual resistive and magnetic behavior of high T$_{c}$ superconductors
(HTSCs) has greatly stimulated the efforts to understand the impact of
disorder, thermal fluctuations and dimensionality on the vortex dynamics\cite
{ref1}. The theory of collective flux creep describes the thermally assisted
motion of vortices in a random potential caused by quenched disorder\cite
{ref2}. Central to the theory of collective flux creep is the elasticity of
the vortex system. Due to the finite values of compress modulus $c_{11}$ and
shear modulus $c_{66}$, when moving from one metastable position to another
one, the vortices have the tendency to jump in bundles with a radius $R_{c}$
in order to balance the elastic energy and the energy related to Lorentz
force. The energy barrier for activation $U_{c}$ increases exponentially
with a decreasing current because $R_{c}$ gets larger and larger. However,
this picture breaks down when {\it plastic motion} of vortices is taken into
account\cite{ref3}. For a 2D vortex lattice, due to the finite energy for
creation of dislocations or dislocation pairs, a plastic motion of vortices
is favorable when the energy barrier exceeds a certain threshold. The
plastic creep sets a cutoff for the exponential growth of $U_{c}$. The
plastic behavior of vortex system has been studied by several groups\cite
{ref4}. Recently, plastic motion of vortices was directly observed on Nb
films with artificially introduced pinning centers by using Lorentz
microscopy\cite{ref5}.

{\it Quantum tunneling} of vortices is expected to replace thermal
activation as the dominant dissipation mechanism when the temperature is low
enough\cite{ref6}. Tunneling of vortices has been observed in transport and
magnetization measurements\cite{ref7,ref8,ref9,ref10,ref11}. It was found
that the relaxation rate of magnetization did not extrapolate to zero when
the temperature went to zero, and this phenomenon was attributed to quantum
tunneling of vortices\cite{ref7}. Variable range hopping of vortices was
observed on ultrathin Pb films\cite{ref9} and YBCO single crystals with
columnar defects\cite{ref10}. Ephron et al.\cite{ref11} observed a
transition from thermally activated to a temperature independent resistance
on Mo$_{43}$Ge$_{57}$ ultrathin films, and this resistance was attributed to
the quantum creep of vortices.

Plasticity of vortex lattice (VL) and tunneling of vortices are two possible
contributions prohibiting the existence of vortex glass state which is
important for future applications. Up to now, they are not well studied and
our understanding to these contributions is still limited. This strongly
motivates further studies of the interplay between elasticity, plasticity
and tunneling of vortices in a 2D vortex system by utilizing suitable
samples. In this paper, we report our study of the vortex behavior in 2D YBa$%
_{2}$Cu$_{3}$O$_{7}$/PrBa$_{2}$Cu$_{3}$O$_{7}$ (YBCO/PBCO) superlattice at
different temperature, magnetic field and excitation current. We observed a
crossover from collective (elastic) creep to plastic creep of vortices when
the elastic limit was reached for a VL, which means a 2D vortex glass can
not exist at finite temperature. A dislocation mediated melting of VL was
observed in the high current limit. A possibility of quantum creep of
vortices at low temperature and thus the non-existence of 2D vortex glass at
T=0 is discussed. Our results provide for the first time a comprehensive
picture for the vortex dynamics within a 2D VL in artificial superlattices.

The c-axis oriented $[YBCO(24 \AA)/PBCO(144 \AA)]_{25}$ superlattice was
fabricated by in-situ magnetron sputtering\cite{ref12}. The X-ray
diffraction showed satellite peak up to the third order, indicating the high
quality of the sample. This film was photolithographically patterned into
strip with a width of 0.1 mm. DC I-V characteristics were recorded with
currents generated by Keithley 238 and voltage detected by Keithley 182.
Resistive transition $\rho _{ac}$(T) was measured by four-terminal ac
locking-in technique with an excitation current of 1 $\mu $A at a frequency
of 17 Hz. The magnetic field was generated by a 15T Oxford superconducting
magnet. During the measurements, the field was kept perpendicular to the
film plane and the temperature stability was better than 10 mK.

Shown in Fig. 1 are the representative I-V characteristics measured at 2 T
with the temperature varied from 2 to 32 K. The I-V curves were taken with a
temperature interval of 0.1 K. Each I-V curve was obtained by averaging over
four individual curves with current applied in positive and negative
directions. Two features are clearly visible: (i) {\it in the high current
limit} ($j>5\times 10^{3}A/cm^{2}$), I-V curves show downward curvature at low temperatures, changing to upward curvature

\begin{figure}[t]
\centerline{
\epsfxsize=3.2 in
\epsfbox{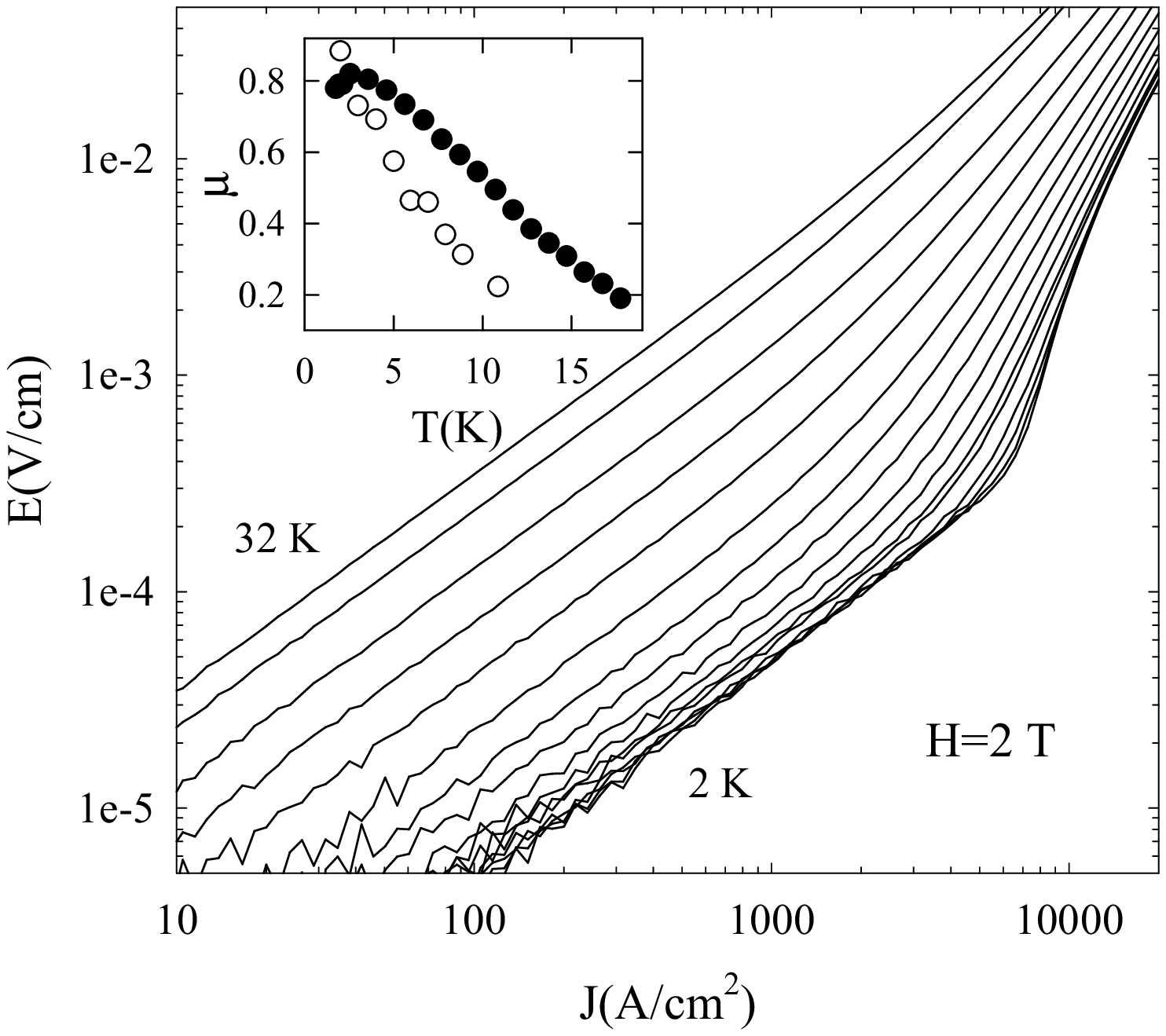}}
\caption{Selected I-V curves at different temperatures ranging from 2 to 32
K with an interval of 2 K. Inset: temperature dependence of $\protect\mu $
at H=0.5 T and 2 T, respectively.}
\label{fig1}
\end{figure}

\hspace{-13pt} at high temperatures; (ii) {\it in the low current limit}, linear I-V curves
were observed for all the temperature region over several order of magnitude
for current and voltage. The linear resistivity $\rho _{L}$ extracted from
the I-V curves are shown in Fig. 2 together with the R-T curves $\rho _{ac}$%
(T) in the Arrhenius plot. We find that $\rho _{L}$ coincides with $\rho
_{ac}$ quite well. In the high temperature part, a thermally activated flux
flow (TAFF) behavior is clearly visible, when the temperature decreases (%
%TCIMACRO{\TEXTsymbol{<}}%
%BeginExpansion
\mbox{$<$}%
%EndExpansion
10 K), the resistivity deviates from the TAFF behavior, and reaches a
constant value in the lower temperature at each field.

Due to the large PBCO layer thickness, YBCO layers are essentially
decoupled, so each YBCO layer can be treated independently. Moreover, each
YBCO layer consists of only two unit cells (24 $\AA$), which is much smaller
than $L_{c}$($\sim 100 \AA)$, the coherence length of vortices
along the c-direction. So we treat the whole vortex system as 2D. To
understand the above I-V characteristics in the whole temperature and
current ranges, we discuss the following 2D vortex dynamics scenarios.

{\it 2D collective creep and vortex melting}- It is expected that in a 2D
disordered superconductor, 2D collective creep of vortices plays an
important role in the dissipation process. 2D collective ceep theory is
applicable when $R_{c}=a_{0}(\varepsilon _{0}d/U_{p})(\xi
/2a_{0})^{2}>a_{0}\approx (\Phi _{0}/B)^{1/2}$, the vortex

\begin{figure}[htb]
\centerline{
\epsfxsize=3.2 in
\epsfbox{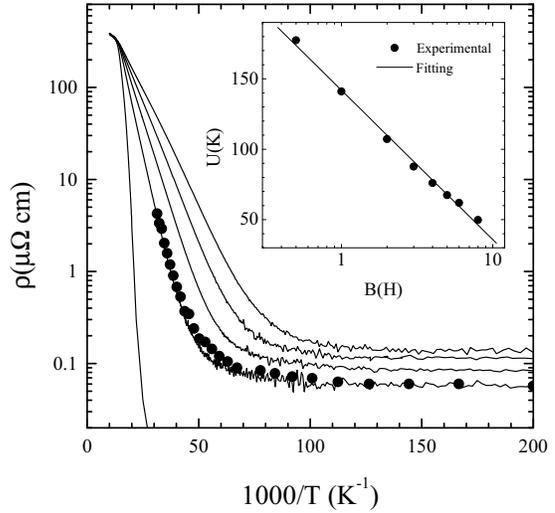}}
\caption{Temperature dependence of $R_{ac}$ at difference applied magnetic
field. The solid lines are experimental data. From left to right: H=0, 2, 4,
6, 8 T. Fitted circles are data points of $R_{L}$ extracted from the linear
parts in Fig.1. Inset: Activation energies from a TAFF fit to $R_{ac}$(T),
filled circles are the obtained data, solid line is a fitting by $U$%
=141.8-105.9$lnB$.}
\label{fig2}
\end{figure}

\hspace{-13pt} 
spacing, where $\varepsilon _{0}=\Phi _{0}^{2}/16\pi ^{2}\lambda ^{2}$ is the energy for a
pancake vortex per unit length, $U_{p}$ the pinning potential, $d$ the
length of the vortex along the field direction, $\xi $ the coherence length 
\cite{ref13}. From Fig. 2 we have $U_{p} \approx 100 K$, with d=24 $\AA$, $\lambda \approx $ 1400 $\AA$, $\xi \approx $ 15 $\AA$, $a_{0}\approx $500 $\AA$, and $R_{c}\sim $5000 $\AA$ $\approx 10a_{0}$. Therefore we
think that 2D collective creep theory is suitable in the present study. When
vortices undergo collective creep, the activation energy is $U(j)\propto
(j/j_{0})^{-\mu }$, where $j_{0}$ is a characteristic current density.
Within the TAFF approximation, the I-V curves caused by the presence of the
energy barrier $U(j)$ is \cite{ref2}

\begin{equation}
E=E_0e^{-\frac{U_0}{kT}(j/j_0)^{-\mu }}\text{ .}  \label{eq1}
\end{equation}

According to Eq. (1), the logarithmic plots of the I-V curves with downward
curvature were fitted by a power law, with $\ln E_{0}$, $(U_{0}/kT)j_{0}^{%
\mu }$, and $\mu $ as fitting parameters. The obtained exponents $\mu $ are
shown in the inset in Fig. 1. The typical values for $\ln E_{0}$ and $%
(U_{0}/kT)j_{0}^{\mu }$ are $\sim $10 and $\sim $1000,
respectively. The $\mu $ values change gradually from 0.8 to 0.2 as the
temperature increases. This result agrees well with that of Dekker et al.
who found the same trend in the varoation of $\mu $ upon changing the
temperature \cite{ref14}. We interpret the change in values of $\mu $ as
follows: as pointed by Vinokur et al.\cite{ref13}, depending on the current
density, temperature and magnetic field, the bundle size will be different
and consequently $\mu $ will take various values. For vortex bundle with
medium size (defined by $j_{lb}\approx j_{c}(R_{c}^{3}/\Lambda
a_{0}^{2})^{8/15}<j<j_{mb}\approx j_{c}(R_{c}/\Lambda )^{8/15}$), an
exponent $\mu $=13/16$\approx 0.8$ is expected. When $j<j_{lb}$, the bundle
size involved in a jump increases, and the exponent $\mu $ changes to 1/2.
We notice that when the magnetic field is fixed, the bundle size is $%
R_{c}\propto \xi ^{2}/\lambda ^{2}U_{p}$. At temperatures not too close to T$%
_{c}$, where $\xi $ and $\lambda $ are nearly constants, $U_{p}$ decreases
with increasing temperature, and therefore $R_{c}$ will increase. As a
result, the bundle size will change from medium to large thus decreasing $%
\mu $ value.

As the temperature increases, dislocation mediated vortex melting occurs
when the following condition is satisfied:\cite{ref15} 
\begin{equation}
Ac_{66}a_{0}^{2}d/kT=4\pi \text{ ,}  \label{eq2}
\end{equation}

\hspace{-13pt} where A $\sim $ 0.4-0.7 is a parameter due to the
renormalization of $c_{66}$ from the defects in the VL and the nonlinear
lattice vibration. We identify the melting temperature as the one at which
the I-V curves at large current show a power law dependence, i.e., $V\propto
I^{\beta }$. For H=2 T, we obtain $V\propto I^{3}$ at T=12.87 K. The shear
modulus $c_{66}$ has been calculated by Brandt\cite{ref16} who showed that 
\begin{equation}
c_{66}=[B_{c2}(t)/4\mu _{0}]b(1-0.29b)(1-b)^{2}\text{ ,}  \label{eq3}
\end{equation}

\hspace{-13pt} where B$_{c}$(t)=B$_{c}$(0)(1-$t^{2}$) is the thermodynamic
critical field, t=T/T$_{c}$ and b=B/B$_{c2}(t)$. Inserting the typical
values of B$_{c}$ and B$_{c2}$ for YBCO with B$_{c}$(0) $\approx $ 1 T and B$%
_{c2}$(0) $\approx $100 T, together with B =2 T, T=12.87 K, and T$_{c}$=31.5
K, we obtain A $\approx $ 0.45, which agrees well with the previously
obtained value\cite{ref4}.

Above the melting temperature, the vortices are in a liquid state pinned by
residual forces. The I-V characeristics can be well described by the classic
Anderson-Kim model with $E=E_{0}\sinh (j/j_{0})$.

We notice that Dekker et al.\cite{ref17} did similar work on ultrathin YBCO
films. In their study, only I-V curves with upward curvatures were observed
and the glass correlation length $\xi _{VG}$ diverges at T=0. They concluded
that a 2D vortex glass could not exist at any temperature T%
%TCIMACRO{\TEXTsymbol{>}}%
%BeginExpansion
\mbox{$>$}%
%EndExpansion
0. In our case, we did observe a crossover from downward to upward
curvatures. This discrepancy could be possibly due to different length scale
involved in the experiments. As we know, when applying a current, a vortex
pair of a characteristic size $L_j=(ckT/j\Phi _0)^{1/2}$ is excited\cite
{ref18}. The current probes the vortex dynamics in a length scale of $L_j$.
With $j\sim 5\times 10^3A/cm^2$, T$\sim $ 10 K, we get $L_j\sim 1000$ $\AA
<R_c$. Therefore, we were probing the collective creep behavior in the high
current limit. While in their case, the length scale was probably larger
than $R_c$, and thus vortex glass behavior was detected.

{\it Plastic motion in the low current limit}- The collective pinning theory
depends critically on the elasticity of vortex system. Due to the finite
energy for the generation of dislocations or dislocation pairs in a 2D VL,
an infinite energy barrier in the elastic limit is not expected. When the
energy barrier reaches the characteristic energy for a small dislocation
pair, it will be cut off and a crossover from elastic to plastic motion
should occur\cite{ref3}. The motion of dislocation pairs can be well
described by TAFF with an activation energy nearly independent of the
current density. Therefore, linear I-V curves are expected in the plastic
region. As a result, there is no vanishing linear resistivity, and thus {\it %
a 2D vortex glass does not exist at any finite temperature. This is exactly
what we observed in our I-V curves}. We see that at a current density about $%
j\sim 5\times 10^{3}A/cm^{2}$, the I-V curves with downward curvature change
to linear at low current densities. This linearity is valid in the range of
more than two orders of magnitude in the current and voltage. The typical
energy for a dislocation pair made up of two edge dislocations separated by $%
a_{0}$ is \cite{ref3} 
\begin{equation}
U_{e}=\frac{\Phi _{0}^{2}d}{16\pi ^{3}\lambda ^{2}(T)}\ln (B_{0}/B)\text{ ,}
\label{eq4}
\end{equation}

With $\lambda \approx $1400 $\AA$, we have $U_{e}\approx $ 500 K. This is
consistent with what we have obtained from a fitting to the I-V curves we
made earlier. With $(U_{0}/kT)j_{0}^{\mu }\sim $ 1000, $j_{0}\sim $ 10$^{5}$
A/cm$^{2}$, we have $U_{0}\sim $ 300 K, which agrees quite well with $U_{e}$%
. This explanation is further supported by the lnB dependence of activation
energies which were extracted from the slopes of the Arrhenius plot as shown
in Fig. 2. So we conclude that {\it in the low current limit, the
dissipation is governed by plastic motion of dislocation pairs in the VL}.

The above picture is consistent with Monte Carlo simulations of the VL
subject to random disorders under a driving force\cite{ref19}. It is found
that at large driving force, the vortex motion is elastic. As the driving
forcing force decreases, the VL becomes defective, and the dissipation is
dominated by plastic motion.

{\it \ Possible quantum tunneling of vortices at low T}- as we can see from
Fig. 2, below T$\sim $10 K, the resistivity gradually deviates from the
activated behavior, showing a level off toward constant values\cite{ref20}.
The resistivity at the plateau grows with the increasing magnetic field. To
get such a plateau within the framework of TAFF would require an activation
energy $U_{0}$ which increases linearly with T, i.e., $U_{0}\propto \alpha $%
T. We are not aware of any mechanism that can provide such kind of an
activation energy. Same resistive plateau in the R-T curves was also
observed by Ephron et al. on Mo$_{43}$Ge$_{57}$ thin films.

Generally, it is believed that quantum tunneling, if it does exist,
dominates the dissipation at temperatures below several Kelvin and that the
resistivity from a quantum creep is nearly independent upon temperature\cite
{ref21,ref22,ref23,ref24}. We are not sure why we observed quantum tunneling
behavior at such a high temperature where the thermal energy of vortices is
quite high. One possibility is a large normal resistivity $\rho _{n}$ and a
small coherence length $\xi $. The tunneling rate $\gamma $ is determined by
the effective Euclidean action $S_{E}^{eff}$ for the tunneling process, $%
\gamma \propto \exp (-S_{E}^{eff}/\hbar )$, and $S_{E}^{eff}/\hbar $ in turn
is proportional to $(\hbar /e^{2})\xi /\rho _{n}$\cite{ref21}. Therefore,
tunneling is favored by a small coherence length $\xi $ and a large normal
state resistivity $\rho _{n}$. Experimentally, Ephron et al. reported that
they were only able to observe such a plateau on thin films with large sheet
resistance, this is consistent with theoretical consideration.

The existing results on the temperature T$_{0}$ at which the dissipation
shows a crossover from thermally activated to a quantum tunneling range from
hundreds mK to tens Kelvin. Ivlev et al.\cite{ref22} showed that, for YBCO
compound, a crossover from thermally activated to quantum tunneling
dissipation happens at T$_{0}\approx 50(j_{c}-j)^{1/2}/j_{c}^{1/2}$ K.
Meanwhile, Stephen\cite{ref23} found that T$_{0}\approx $2 K.
Experimentally, a crossover temperature ranging from 3.5 K to 10 K was
reported for TlBaCaCuO single crystals\cite{ref8}. When vortices are
localized by columnar defects, a crossover temperature from collective creep
to quantum creep (variable range hopping) as high as 35 K was reported in
the high current limit ($\sim $10$^{6}$A/cm$^{2}$)\cite{ref10}. It is worth
noting that Blatter et al.\cite{ref21} found the suppression of the
tunneling rate for increasing field. In this case the interaction between
voritces increases, the viscocity grows and more vortices are participating
in a single tunneling event. This is consistent with our observations. We
found that the temperature, at which the resistivity starts to deviate from
TAFF behavior, decreases with the increasing magnetic field.

\begin{figure}[t]
\centerline{
\epsfxsize=3.5 in
\epsfbox{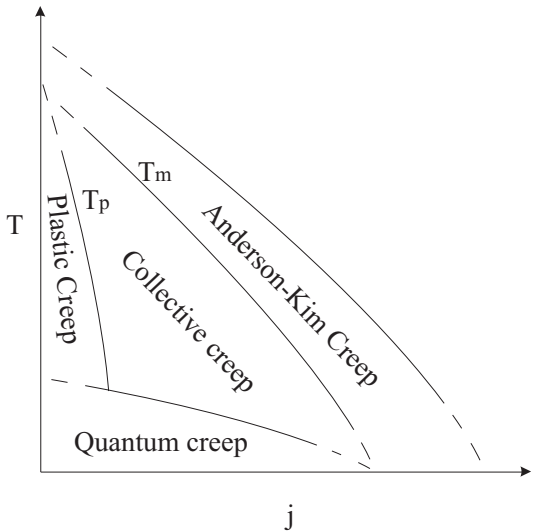}}
\caption{A schematic j-T phase diagram for a 2D VL at a certain magnetic
field. The line T$_{m}$ is determined by Eq. (2), and line T$_{p}$ by Eq.
(4). In the quantum creep and plastic creep regions, the linear resistivity
does not vanish. The collective creep and Anderson-Kim creep regions are
characterized by I-V curves with $E\propto e^{j^{-\mu }}$ and $E\propto \sinh (j/j_{0})$, respectively.}
\label{fig3}
\end{figure}

Therefore, our results suggest the possibility of tunneling of dislocation
pairs which are eqivalent to a vacancy or an interstitial in a VL, and {\it %
a 2D vortex glass can not exist even at T=0 K}. At present, due to the lack
of a suitable theoretical model, we are not able to do a quantitative
analysis of the appearance of the puzzling plateau in the low T region.
Further theoretical and experimental work at that subject is greatly
encouraged.

The main conclusions of this work are summarized in fig. 3. We found that
the vortex dynamics is essentially length scale dependent. In the high
current limit, 2D collective creep was observed which is elastic in nature
and characterized by an energy barrier with an exponential growth with
current density. The 2D VL undergoes a dislocation mediated melting at a
temperature T$_{m}$. In the low current limit, the exponential growth of
energy barrier is cut off by the generation of dislocation pairs in the VL
and their subsequent motion or quantum tunneling. The linear resistivity at
low temperatures ($\lesssim $10 K) is possibly governed by quantum tunneling
of pancake vortices. With increasing Lorentz force, the driven VL recovers
its elasticity at high current density.

This research has been supported by the Belgian IUAP and Flemish GOA, FWO
and Bilateral Flemish-Chinese (BIL 97/35) programs.


\begin{references}
\bibitem[{*}]{byline}  Present Address: Research Institute for Engineering
Materials, Tokyo Institute of Technology, Nagatsuta 4259, Yokohama 226,
Japan.

\bibitem{ref1}  G. Blatter et al., Rev. Mod. Phys. {\bf 66}, 1125(1994).

\bibitem{ref2}  M.V. Feigel\'{}man et al., Phys. Rev. Lett. {\bf 63},
2303(1989); K.H. Fischer and T. Nattermann, Phys. Rev. {\bf B 43},
10372(1991).

\bibitem{ref3}  M. V. Feigel\'{}man, V. B. Geshkenbein, and A.I. Larkin,
Physica {\bf C 167}, 177(1989); V.M. Vinokur, P.H. Kes and A.E. Koshelev,
Physica {\bf C 168}, 29(1990).

\bibitem{ref4}  P. Berghuis et al., Phys. Rev. Lett. {\bf 65}, 2583(1990);
X.G. Qiu et al., Phys. Rev. {\bf B 48}, 16180(1993); Y. Abulafia et al.,
Phys. Rev. Lett. {\bf 77}, 1596(1996).

\bibitem{ref5}  T. Matsuda et al., Science 271, 1393(1996); K. Harada et
al., Science 274, 1167(1996).

\bibitem{ref6}  H. Grabert and U. Weiss, Phys. Rev. Lett. {\bf 53},
1787(1984).

\bibitem{ref7}  A.C. Mota et al., Physica {\bf C 185-189}, 343(1991); R.
Griessen et al., Physica {\bf C 185-189}, 337(1991); A. Gerber and J.J.M.
Franse, Phys. Rev. Lett. {\bf 71}, 1895(1993).

\bibitem{ref8}  X. Zhang et al, Physica {\bf C 235-240}, 2957(1994).

\bibitem{ref9}  Liu et al., Phys. Rev. Lett. {\bf 68}, 2224(1992).

\bibitem{ref10}  J.R. Thompson et al., Phys. Rev. Lett. {\bf 78}, 3181(1997).

\bibitem{ref11}  D. Ephron et al., Phys. Rev. Lett. {\bf 76}, 1529(1996).

\bibitem{ref12}  G. Jakob et al., Appl. Phys. Lett. {\bf 59}, 1626(1991).

\bibitem{ref13}  V.M. Vinokur, P.H. Kes and A.E. Koch, Physica {\bf C 248},
179(1995)

\bibitem{ref14}  C. Dekker, W. Eidelloth and R.H. Koch, Phys. Rev. Lett. 
{\bf 68}, 2247(1992).

\bibitem{ref15}  B.A. Huberman and S. Doniach, Phys. Rev. Lett. {\bf 43},
950(1979).

\bibitem{ref16}  E.H. Brandt, Phys. Status Solidi B {\bf 77}, 551(1976).

\bibitem{ref17}  C. Dekker et al., Phys. Rev. Lett. {\bf 69}, 2717(1992).

\bibitem{ref18}  D.S. Fisher, M.P.A. Fisher and D.A. Huse, Phys. Rev. {\bf B
43}, 130(1991).

\bibitem{ref19}  H.J. Hensen et al., Phys. Rev. {\bf B 38}, 9235(1988); A.C.
Shi and A.J. Berlinsky, Phys. Rev. Lett. {\bf 67}, 1926(1991).

\bibitem{ref20}  An explanation by experimental error is not favored, since
the resistivity at the plateau corresponding to a voltage of 0.1 $\mu $V, at
least two order higher than the precision of our instruments. The
resistivity at the plateau corresponds to a sheet resistance of 0.1 $\Omega $%
, about one order lower than that reported by Ephron et al.

\bibitem{ref21}  G. Blatter, V. B. Geshkenbein and V.M. Vinokur, Phys. Rev.
Lett. {\bf 66}, 3297(1991); Phys. Rev. {\bf B 47}, 2725(1993).

\bibitem{ref22}  B.I. Ivlev, Yu.N. Ovchinnikov, and R.S. Thompson, Phys.
Rev. {\bf B 44}, 7023(1993).

\bibitem{ref23}  M. J. Stephen, Phys. Rev. Lett. {\bf 72}, 1534(1994);

\bibitem{ref24}  M.W. Gaber and B.N. Achar, Phys. Rev. {\bf B 52},
1314(1995).
\end{references}
\end{document}